# Evolution of Porosity in Suspension Thermal Sprayed YSZ Thermal Barrier Coatings through Neutron Scattering and Image Analysis Techniques

Daniel Tejero-Martin[1], Mingwen Bai[2], Jitendra Mata[3], Tanvir Hussain[1*]

[1] *Faculty of Engineering, University of Nottingham, University Park, Nottingham, NG7 2RD, UK*

[2] *Institute for Future Transport and Cities, Coventry University, Priory St, Coventry, CV1 5FB, UK*

[3] *Australia's Nuclear Science and Technology Organisation (ANSTO), New Illawarra Rd, Lucas Heights NSW 2234, Australia*

* tanvir.hussain@nottingham.ac.uk; +44 115 951 3795

**Abstract**

Porosity is a key parameter on thermal barrier coatings, directly influencing thermal conductivity and strain tolerance. Suspension high velocity oxy-fuel (SHVOF) thermal spraying enables the use of sub-micron particles as feedstock, increasing control over porosity and introducing nano-sized pores; these fine-scale porosities being challenging to measure. Neutron scattering represents a non-destructive technique, capable of studying porosity with a pore size range of 1 nm to 10 µm, thanks to the combination of small-angle (SANS) and ultra-small-angle neutron scattering (USANS) techniques. Image Analysis (IA) on digital images allows for the study of porosity with a size above ~100 nm. In this work, two yttria-stabilised zirconia ($ZrO_2$ with 8 wt.% $Y_2O_3$) suspensions were sprayed and heat treated at 1100, 1200, 1300 and 1400 °C for 72 h. For the first time in SHVOF 8YSZ, pore size distribution, total porosity and pore morphology were studied using SANS & USANS and IA to determine the effects of heat treatment. X-ray diffraction and micro-hardness measurements were performed to study the phase transformation, and its effects on the mechanical properties of the coating. The results show an abundant presence of nano-pores in the as-sprayed coatings, which are eliminated during heat treatment at 1100 °C; a transition from inter-splat lamellar pores to globular pores and the appearance of micro-cracks along with the accumulation of micro-strains associated with the phase transformation from the initial metastable tetragonal into tetragonal and cubic phases at 1200 °C. The phase transformation was completed at 1400 °C, with no presence of monoclinic phase.

**Keywords:** neutron scattering; image analysis; porosity; suspension thermal spray; YSZ

## 1. Introduction

Yttria-stabilised zirconia (YSZ) currently represents the most used material within thermal barrier coatings (TBC) for gas turbine engine components [1,2]. The main goal of a successful TBC is to provide insulation to the substrate underneath, effectively reducing the experienced temperature and avoiding component degradation due to excessive surface temperature. 8YSZ presents a low thermal conductivity (0.7 – 1.4 W/mK), good thermal stability (within its temperature application range) and a coefficient of thermal expansion (CTE) close to that of the commonly used Ni-based superalloys (~11 × $10^{-6}$ $K^{-1}$ for 8YSZ coatings and ~14 × $10^{-6}$ $K^{-1}$ for the Ni superalloys) [3], making it a sound choice for a TBC topcoat. In addition to its inherent properties, a successful 8YSZ TBC must present a favourable microstructure, in terms of porosity and micro-cracking [4]. For instance, coatings produced using electron-beam physical vapour deposition (EB-PVD) present columnar structure, leading to higher thermal conductivity values, but improved strain tolerance and thermal shock resistance. When atmospheric plasma spraying (APS) is used, the coatings present a splat-based layered structure, with the advantage of a reduced thermal conductivity and production costs when compared to EB-PVD.

Even within APS deposited coatings, the variation of the amount of total porosity and pore size distribution has an effect on the thermomechanical properties of the system, directly affecting properties such as the hardness [5] and the accumulation of thermal stresses [6]. The pore size in a typical APS deposited coating has radii between 0.02 and 1 µm [2]. An increase in the total porosity correlates with a reduction in the density of vertical cracks [7] (also known as segmentation cracks, responsible for an increase in thermal conductivity), a reduction of the residual compression stresses [8]. Thermal insulation in YSZ, at service temperatures of ~1200 °C, is improved by phonon scattering at inhomogeneities (grain boundaries, gas-filled pores and vacancies) [9]. More porosity provides more scattering sites, effectively lowering the thermal conductivity of the coating. It is clear then, that a more precise control over the porosity present within the as-deposited coating is desirable when manufacturing TBCs with enhanced properties and performance. Since the porosity in thermal sprayed coatings is strongly correlated to the size distribution of the feedstock particles, mostly due to unmelted feedstock particles and gaps in between adjacent splats [10], there has been a growing interest for sub-micron feedstock particle distribution. Such a reduction in size allows finer grains, higher strength and durability, reduced porosity sizes [11] and enhances the thermal and mechanical properties of the coating [6,12].

Despite the potential shown by the use of nanostructured YSZ powders, APS deposited coatings present a lower limit on the feedstock particle size of around 10 – 100 µm [13,14] to ensure adequate flowability. Aiming to circumvent this issue, suspension thermal spray was developed, where the feedstock material is presented in sub-micron size and dispersed in a liquid medium (generally water or ethanol) allowing the use of particles with a smaller size. Such approach has led to the development of novel thermal spraying deposition techniques such as suspension plasma spray (SPS) or suspension high velocity oxy-fuel spray (SHVOF). Suspension thermal spray techniques have been successfully applied to the deposition of YSZ coatings [15], with special focus on the effects on the physical and thermal properties of the deposited coatings [16]. SHVOF thermal spray has been reported in the past as the deposition technique for 8YSZ [11], with higher thermal conductivities than coatings produced using SPS due to the presence of vertical cracks. Nevertheless, we demonstrated that SHVOF thermal spray can be used to deposit crack-free coatings, with the porosity being variable depending on the spraying parameters and suspension medium used [17]. Despite these promising results, no thorough investigation on the porosity of SHVOF thermal sprayed 8YSZ coatings has been reported, particularly aiming at the study of nano-porosity and the microstructural evolution at service temperatures.

Due to the nature of the environment to which TBCs are exposed during service, where temperatures of 1200 °C are expected, the deposited coating will experience several heat-induced phenomena.

Microstructural changes are to be expected, such as coalescence of pores (coarsening) or closure of pores (sintering) as well as the appearance of micro-cracks, having a direct impact on the thermal conductivity of the coating. The importance of a deep understanding of the evolution that the porosity on thermal sprayed TBCs undergoes during heat treatment is evidenced by the abundant literature in the topic [9,16–23]. Despite the recognised importance and the essential role that porosity plays in TBCs, accurate measurement of the porosity remains a challenging task that should be approached carefully. A plethora of techniques have been developed over the years to measure porosity, with each one of them having its own set of advantages and disadvantages, which should be considered when performing the measurements. One of the key factors is the measurable pore size range of the technique. Scanning electron microscopy (SEM) image analysis is a commonly used technique for the determination of the porosity on thermal sprayed coatings [24], but it is limited to features with a size above ~100 nm [25,26]. Such size limit is inadequate for the study of suspension thermal sprayed coatings, containing pores with radii below 10 nm [27]. X-ray/neutron scattering techniques have already been reported to be effective for the measurement of porosity, both in different materials as here presented [28,29] and in YSZ coatings [30–34]. Such techniques offer access to a wide range of size pores, particularly if the ultra-small-angle variations are considered, allowing pores with a radii ~1 nm to be studied [35,36].

In this work, a comprehensive study of the evolution of porosity on SHVOF thermal sprayed 8YSZ coatings during heat treatment at 1100 °C, 1200 °C, 1300 °C and 1400 °C for 72 h is presented. Pore size distribution and total porosity were measured using neutron scattering techniques (SANS and USANS) and IA. To further understand the effects of porosity, phase composition and micro-hardness were measured and correlated to the pore size distribution and total porosity measured.

## 2. Experimental methods

### 2.1. Materials and coating deposition

Two commercially available 8YSZ ethanol-based suspension were used in this study, one supplied by Oerlikon (Pfäffikon, Switzerland) being referred to as O-YSZ here, and the other one supplied by Treibacher Industrie AG, (Althofen, Austria), being referred to as T-YSZ. To avoid differences arising from the yttria content or solid content of the suspension, both suspensions had an 8 wt.% yttria content and the solid content of the suspension was 25 wt.% as supplied by the manufacturers. Particle size distribution (PSD) for both suspensions had very similar values, as stated by the manufacturers. O-YSZ had a $d_{90}$ value of 0.8 – 2.0 μm, $d_{50}$ value of 0.3 – 1.0 μm and $d_{10}$ value of ~0.1 μm. T-YSZ had a $d_{90}$ value of 1.32 μm, $d_{50}$ value of 0.60 μm and $d_{10}$ value of 0.26 μm.

The coatings were deposited using a modified GTV TopGun HVOF thermal spray system with direct injection of suspension at the centre of the gas mixing block. The injector had a diameter of 0.3 mm, the length of the combustion chamber was 22 mm and a 110 mm long barrel nozzle was used. A detailed description of the setup can be found elsewhere [12]. The suspensions were homogenised for at least 2 hours prior to the spraying using a roller mixer and sealed containers to avoid evaporation of the ethanol. Further homogenisation was provided through the spraying via a mechanical stirrer in the pressurised vessel. Mild steel substrates with dimensions of 60 × 25 × 2 mm were used, being attached to a carousel with a diameter of 260 mm, a rotation speed of 73 rpm and the gun traverse speed being set to 5 mm/s, corresponding to a surface speed of 1 mm/s. 40 passes of the gun were completed to deposit the coatings. The hydrogen flow rate used for the deposition of both suspensions was 700 l/min and the oxygen flow rate was 300 l/min, providing a theoretical flame power of 99 kW. Before spraying, the substrates were subjected to grit blasting with a blast cleaner (Guyson, UK) using fine F100 brown alumina (0.125 – 0.149 mm) particles at 3 bar. Following grit blasting the substrates were cleaned in industrial methylated spirit using an ultrasonic bath for up to 10 min and dried with compressed air.

Free-standing coatings were produced submerging the coated mild steel substrate in HCl 37 wt.% for 2 – 4 hours, until the coating detached from the substrate. The produced free-standing coatings were subjected to heat treatment using an Elite Thermal Systems Ltd. (Leicestershire, UK) BRF14/5 box furnace at temperatures of 1100 °C, 1200 °C, 1300 °C and 1400 °C, with a heating rate of 10 °C/min in air for 72 h. Once the heat treatments were finished, the samples were allowed to cool down inside the furnace to prevent drastic temperature drops that would induce stresses.

### 2.2. Neutron scattering

SANS and USANS measurements were performed on free-standing samples using the QUOKKA [37] and KOOKABURRA [38] instruments, respectively, at the Australia's Nuclear Science and Technology Organisation (ANSTO) using the OPAL reactor (Sydney, Australia). For the SANS measurements, an incident neutron beam with wavelengths of 5 Å and 8.1 Å (Δλ/λ = 10%) were used, with sample-to-detector distance of 12 m, 20 m (with focussing lens optics in the latter case). USANS measurements were carried out using an incident neutron beam with a wavelength of λSi(311) = 2.37 Å (Δλ/λ = 4%). The acquired data was corrected for sample transmission, empty cell scattering, detector sensitivity and background scattering. The corrected data was scaled to absolute intensities by comparison to empty beam flux using a package of macros in Igor software (Wavemetric, USA) modified to accept data files from QUOKKA and KOOKABURRA. The SANS and USANS data were then desmeared and merged. Raw data included the uncertainty of each data point, showing typical values of ~0.5 – 3.5 % for low scattering angles and ~10 – 20% for high scattering angles. Due to the data analysis procedure, uncertainty in the raw data did not translate into uncertainty on the obtained volume distribution or total porosity. More details of the neutron scattering techniques and the consequent data analysis is presented in Appendix A.

### 2.3. Material characterisation

Cross-sections of the coatings were prepared by cold mounting a free-standing coating using EpoFix resin and hardener (Struers, Denmark) and cure it for 24 hours. The mounted free-standing was then ground and polished to a 1 μm finish using SiC grinding papers (Buehler, Germany). For the SEM images a Quanta 600 (FEI Europe, Netherlands) scanning electron microscope was used to image the cross-section of the free-standing coatings using backscattered electron (BSE) mode. Imaging parameters were kept constant, with an accelerating voltage of 15 kV, a spot size of 3 and a working distance of 10 mm. A magnification of 1500× was used for the IA, taking 5 images on different regions of the coating which contained representative porosity. The magnification was chosen as a balance between the need to encompass sufficient microstructural features, allowing for valid averaging results, while obtaining high enough resolution to capture detailed features. Areas with unusual large features, such as cracks, were avoided [24].

Phase determination was carried out using a D8 Advance Da Vinci diffractometer (Bruker, Germany) with Cu cathode (wavelength of 1.5406 Å) using transmission mode on free-standing coatings. The angular range investigated was from 20° to 90° for the complete spectra, and from 72° to 76° for detailed measurements. Step size was set to 0.02° and dwell time was 1 s for all the measurements. Rietveld refinement (TOPAS v4.2 software package) was used to determine the $c/a\sqrt{2}$ parameter, where $c$ and $a$ are the unit cell dimensions. Quantitative Rietveld refinement was employed to determine the quantity of each phase ($t$-, $t'$- and $c$-phase), and principles of whole powder pattern modelling (WPPM) were used for crystallite size and micro-strain calculations [39–41]. Micro-hardness measurements were performed on the cold mounted, polished cross-sections using a Vickers micro-hardness indenter (Buehler, USA) using a load of 200 gf and a dwell time of 10 s. Five indentation, approximately in the middle of the cross section and sufficiently spaced in between them were measured to calculate the micro-hardness value, being the error calculated as the standard deviation.

### 2.4. Image analysis

Five SEM-BSE images were taken at different areas of the centre of cross-section of the coating. All five images per sample were analysed using the open source software ImageJ with the image processing package "Fiji" [42]. A macro was written for the batch preparation of the images. The macro was written so that first it would set the appropriate scale bar and remove the data bar from the image, then setting a specified threshold converting the image into a binary black and white map. It should be noted that the threshold was manually chosen for each set of images after visual inspection of the SEM-BSE images in order to maximise the porosity detected while maintaining noise to a minimum, as indicated by the corresponding ASTM E2109-01 standard [43]. The white and black map was then analysed to measure first the area covered with pores with a size above a determined minimum size to provide the total porosity of the coating. This cut-off minimum size was chosen to filter out single pixel instances, associated with noise rather than physical pores. Then, the area of each individual pore detected in the white and black map was measured and the distribution of all the values was calculated within ImageJ manually setting the size of the bin to 0.008 $\mu m^2$. The frequency for each bin was averaged using the value from each of the five SEM-BSE images, being the standard deviation calculated as well. From the area, the radius of the equivalent sphere was calculated. Although not all the pores are expected to be perfect spheres, this criterion was chosen since a similar assumption was made in the analysis of the neutron scattering data.

## 3. Results

### 3.1. Porosity measurements

As outlined in the Introduction, the pore size measurable is determined by the technique chosen to measure the porosity. The main purpose of this work is to present a comprehensive study of the evolution of the porosity over a wide pore radii range using neutron scattering and IA. It is evident from the results that the combination of SANS and USANS allows the study of pores with a radius between ~1 nm and ~10 µm, effectively encompassing the nano-sized features expected from suspension thermal spray and traditional micron-sized features. The overall response to a heat treatment process was in the two samples a striking modification of both the pore volume distribution profile and the total porosity present. Figure 1 presents the information extracted from the neutron scattering data after modelling, where the volume distribution of the porosity present within each coating is shown.

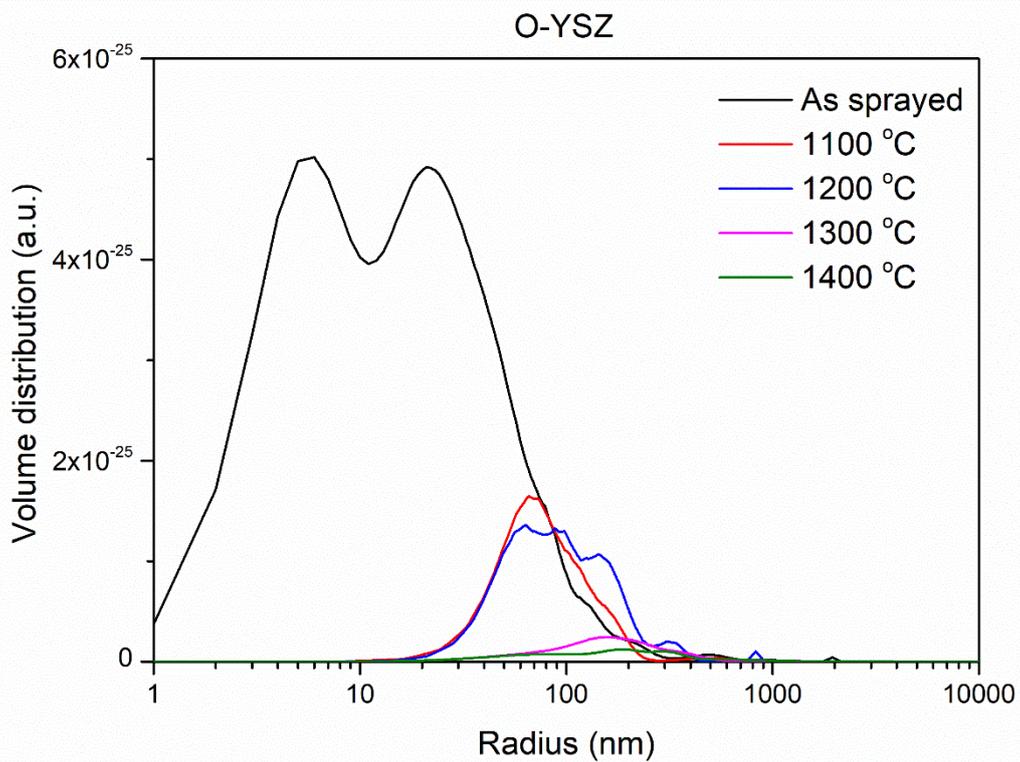

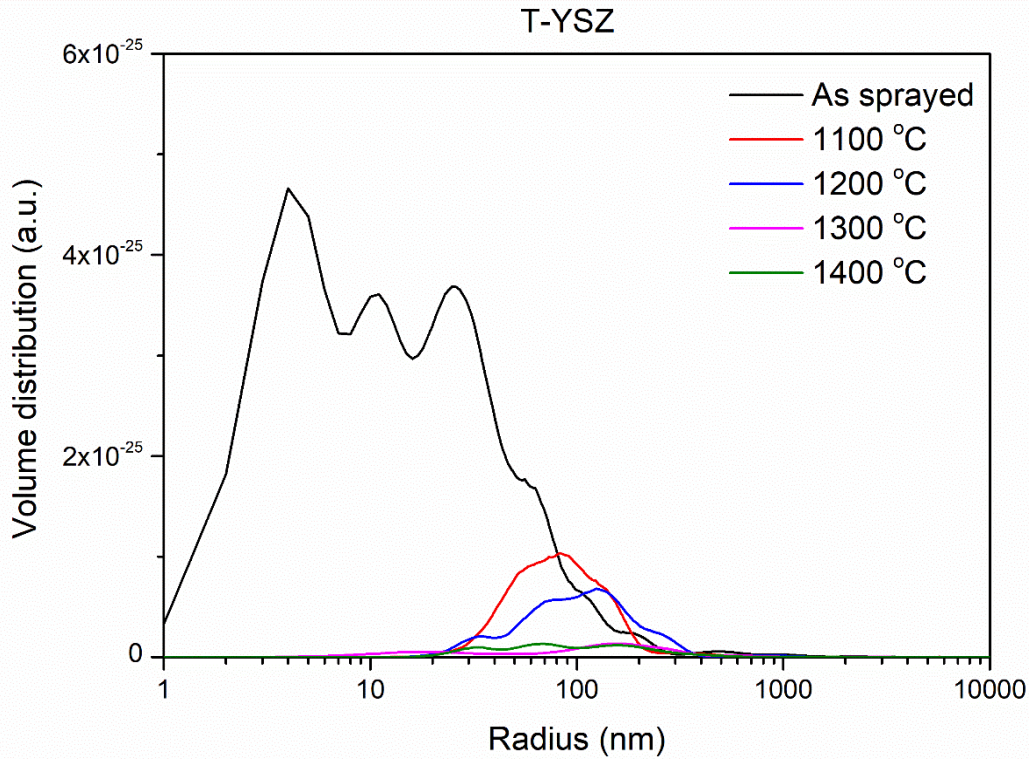

*Figure 1: Volume distribution of the porosity of O-YSZ and T-YSZ coatings in as-sprayed condition and after 72 h at various temperatures, measured using SANS and USANS. The results show that porosity with radii below 20 nm is greatly reduced once heat treatment is performed, with an overall reduction in the total porosity as the heat treatment temperature is increased*

One immediate observation that can be made from Figure 1 is that for both O-YSZ and T-YSZ coatings, there is a noticeable change between the as-sprayed samples and the samples heat treated at 1100 °C. Both as-sprayed samples have virtually all their porosity located with radii < 300 nm, having most of the pores radius < 100 nm. As it was discussed previously, suspension thermal sprayed coatings present nano-size porosity with radius below 100 nm, outside of the accessible range of IA, making SANS and USANS the appropriate technique. Once a heat treatment is performed, even at the lowest temperature of 1100 °C, the nano-pores with a radius below 10 nm effectively disappears in both coatings. This reduction in nano-sized porosity is accompanied by an overall reduction of the total porosity measured via neutron scattering. Data in Figure 1 suggest that both coatings behave slightly different when heat treated at 1100 °C and 1200 °C. In the case of O-YSZ a small reduction in the porosity with radius below 100 nm can be seen, as well as the appearance of porosity with radius above 100 nm. Additionally, there is the appearance of a small population of pores with radius ~900 nm. Regarding the T-YSZ coating, there is a considerable larger reduction in porosity with radius below 100 nm, with only a minor increase in porosity with a radius above ~200 nm. No signs of populations of pores at larger radii could be observed. The behaviour observed for both samples when heat treated at 1300 °C is fairly similar, with an overall reduction in the total porosity measured and flatter pore volume distributions, showing pore sizes more evenly distributed instead of clearly grouped in populations. The changes related to the samples heat treated at 1400 °C, both O-YSZ and T-YSZ, are less obvious from the pore volume distribution plot, presenting a similar profile to the corresponding samples heat treated at 1300 °C.

SEM-BSE images of polished cross-section of the coatings were taken and analysed to measure the porosity using IA. This technique allows not only for the determination of the total porosity within the coating, but to investigate the microstructure of the coatings in the as-sprayed condition as well as after heat treatment. High magnification images of the as-sprayed and after heat treatment at 1400 °C cross section of the O-YSZ and T-YSZ coatings can be observed in Figure 2.

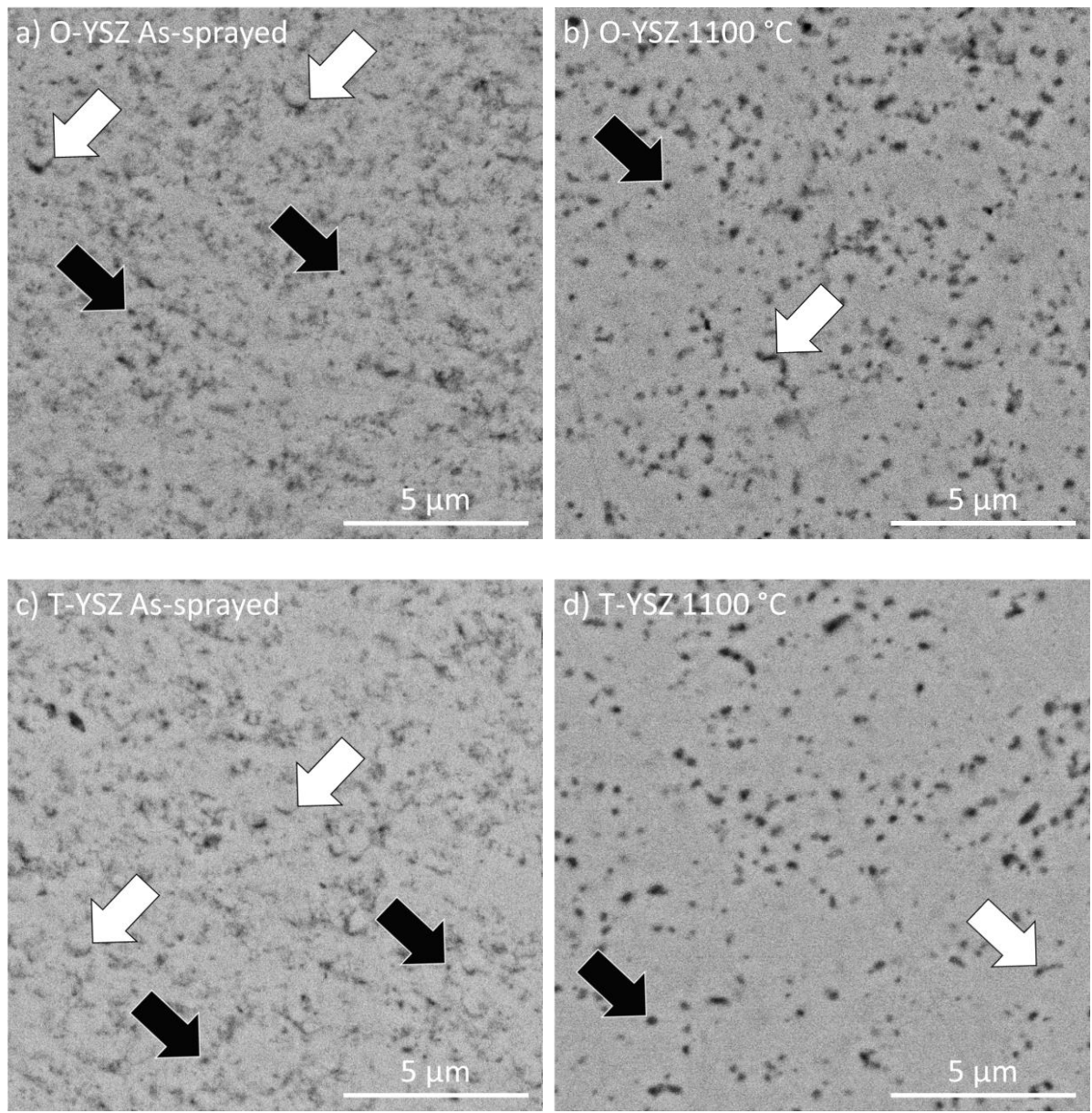

Figure 2: High magnification SEM-BSE images of the cross section of the O-YSZ (images a and b) and T-YSZ (images c and d) in the as-sprayed condition and heat treated at 1100 °C for 72 h. Black arrows mark globular porosity, while white arrows mark non-globular porosity. An overall reduction in the porosity and a transformation from non-globular to globular pores can be appreciated

As it can be seen, for both coatings the samples in the as-sprayed condition (Figure 2a and c) present higher level of porosity than the heat-treated coatings, with abundance of inter-splat porosity. The morphology of the pores evolves into more globular structures as the heat treatment is conducted, being the effect more predominant the higher the heat treatment temperature is (not shown here). Additionally, it can be easily appreciated that the overall level of porosity is reduced as the heat treatment is conducted. An increase of the heat treatment temperature further continued this process, seeing a reduction in the inter-splat porosity with a transformation into spherical pores, and an overall reduction of the total porosity observed.

SEM-BSE cross-section images of the coatings were analysed to determine the apparent pore size distribution. The data can be used for comparison with the pore distribution profiles obtained using neutron scattering, as it is shown in Figure 3.

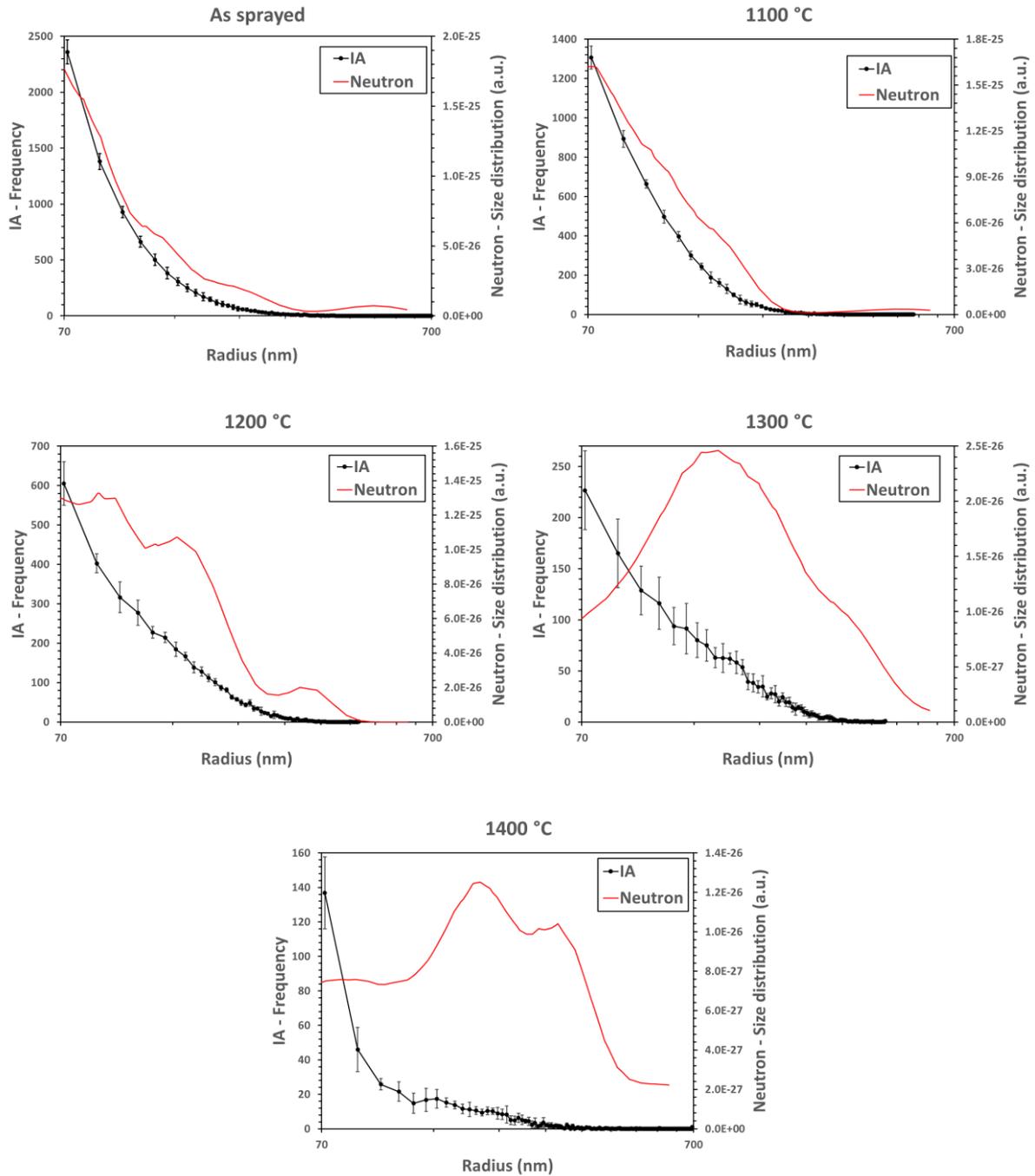

Figure 3: Size distribution of the porosity measured using SEM-BSE cross-section images (black) and neutron scattering (red). The results show good agreement on pore size distribution for the as-sprayed samples, with diverging profiles as the heat treatment temperature is increased, inducing microstructure changes not captured by IA

Pore size distribution data show good agreement within the available range (70 – 700 nm) between IA and neutron scattering for the as-sprayed samples. Once the samples are heat treated, microstructural changes take place. The differences between IA and neutron scattering become more apparent as the temperature is increased, particularly at 1300 °C and 1400 °C, where the local maximum does not match for both techniques.

Neutron scattering and IA also provide information on the total porosity of the coating. The results from both techniques can be seen in Figure 4, where the total porosity, measured with both neutron scattering and IA techniques, is presented.

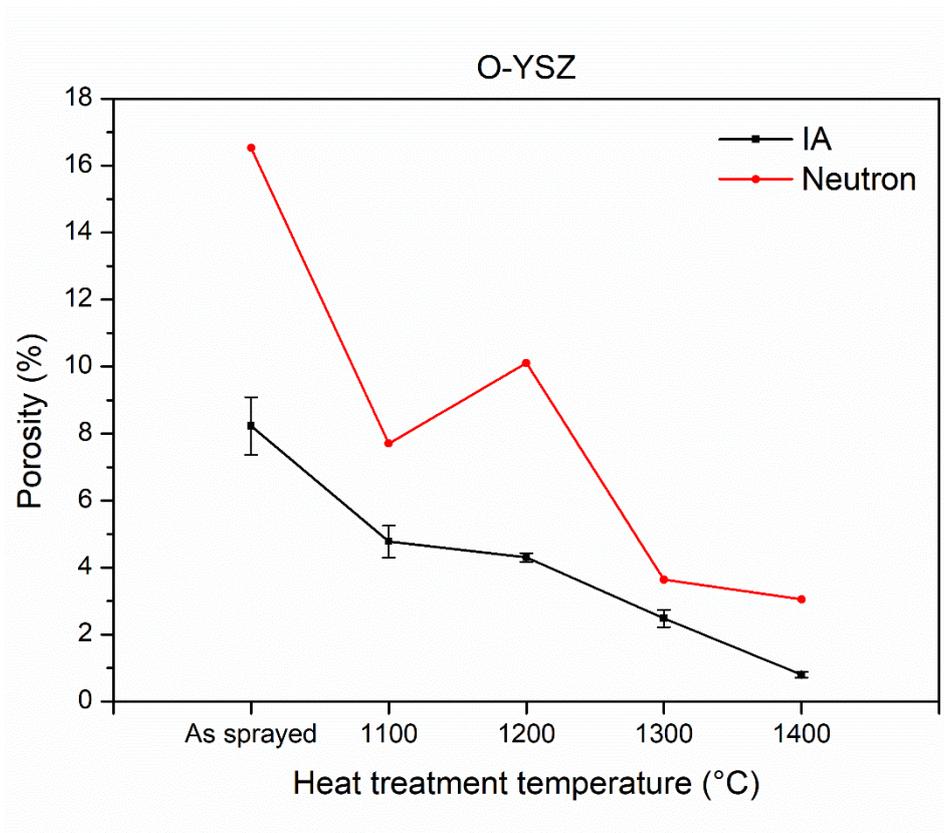

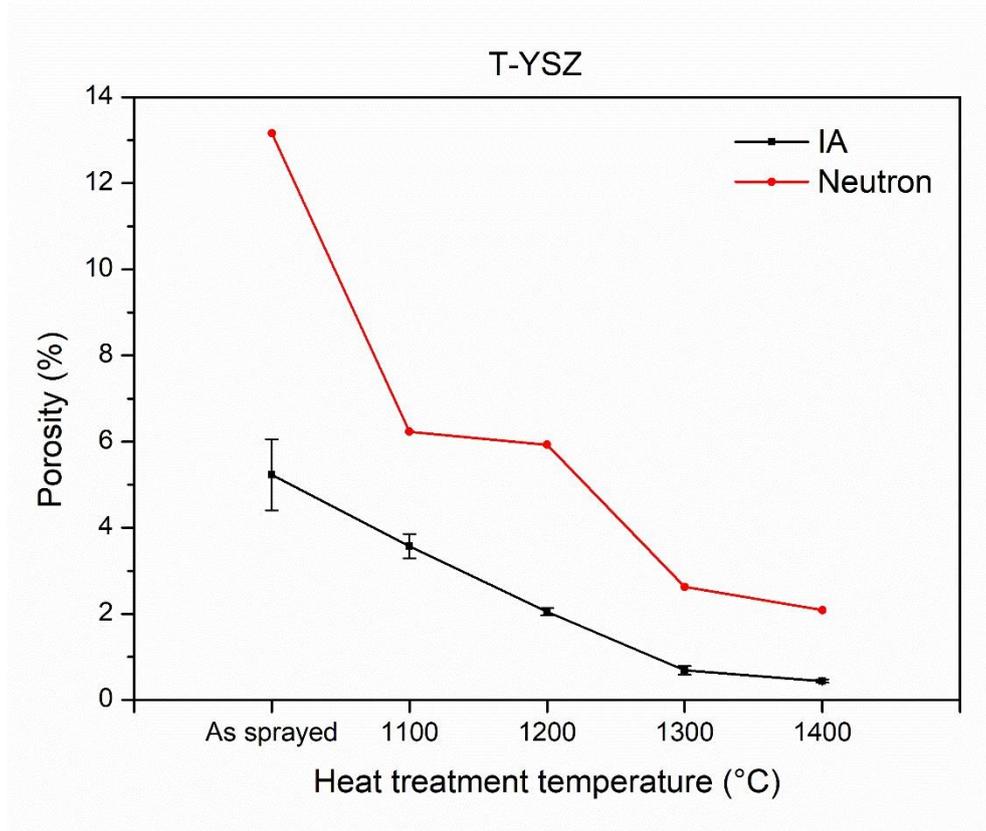

*Figure 4: Total porosity in the as-sprayed as well as heat-treated samples for O-YSZ and T-YSZ coatings, measured using image analysis (black) and neutron scattering (red). The results show an overall reduction in porosity as the heat treatment temperature is increased. Neutron scattering measurements produce a higher porosity when compare to IA measurements*

The data in Figure 4 indicates that, for both coatings, neutron scattering measures a higher total porosity when compared to IA. This difference is particularly notable in the as-sprayed samples. Regarding the O-YSZ coatings, there is a sharp increase in the porosity measured using neutron scattering in the sample heat-treated at 1200 °C, contrary to the observed tendency of a reduction in the total porosity as the heat treatment temperature is increased. When measured using IA, the porosity decreases between 1100 °C and 1200 °C 10%, a much lower value compared to the 40 – 60% reduction experienced for the rest of the temperatures. The data seems to indicate that there is an unknown phenomenon in sample O-YSZ 1200 °C that neutron scattering data is reflecting, but IA is not fully capturing. As for the T-YSZ 1200 °C sample, there could be signs of a similar phenomenon, although neutron scattering does not show an increased porosity. Regardless, the reduction in porosity at 1200 °C is 5%, being notably lower than the 20 – 50% reduction in the rest of temperatures. IA data does not show any distinctive feature, suggesting that if T-YSZ is experiencing a similar process than O-YSZ, its magnitude is lower, and in any case below the detection limit of the IA technique.

3.2. Phase composition

To understand the different behaviour of the coatings when heat treated at 1200 °C, the phase content was investigated using XRD measurements. It is a generally well-known fact that 8YSZ undergoes phase transformation when exposed to high temperatures. During prolonged heat treatment above 1200 °C [20,44], the initial metastable tetragonal phase (*t'-YSZ*) decomposes into yttria-lean tetragonal phase (*t-YSZ*) and yttria-rich cubic phase (*c-YSZ*). If the decomposition process continues, the coating will reach a state where monoclinic (*m-YSZ*) and cubic are the only two phases present. The phase transformation process is as follows:

$$t'\text{-}YSZ \rightarrow t\text{-}YSZ + c\text{-}YSZ \rightarrow m\text{-}YSZ + c\text{-}YSZ$$

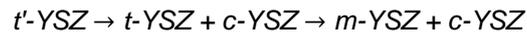

The XRD spectra of the as-sprayed and heat-treated coatings are presented in Figure 5. A scan over the entire range (20° - 90°) is presented for both coatings, with a more detailed scan in the relevant range (72° - 76°) for precise phase identification being shown as well. The cubic phase is generally considered to be detrimental due to its lower fracture toughness when compared to the metastable tetragonal phase [3]. The transformation from tetragonal to monoclinic should be avoided as well, as it implies a volume expansion of 4% [45] that can induce failure in the coating.

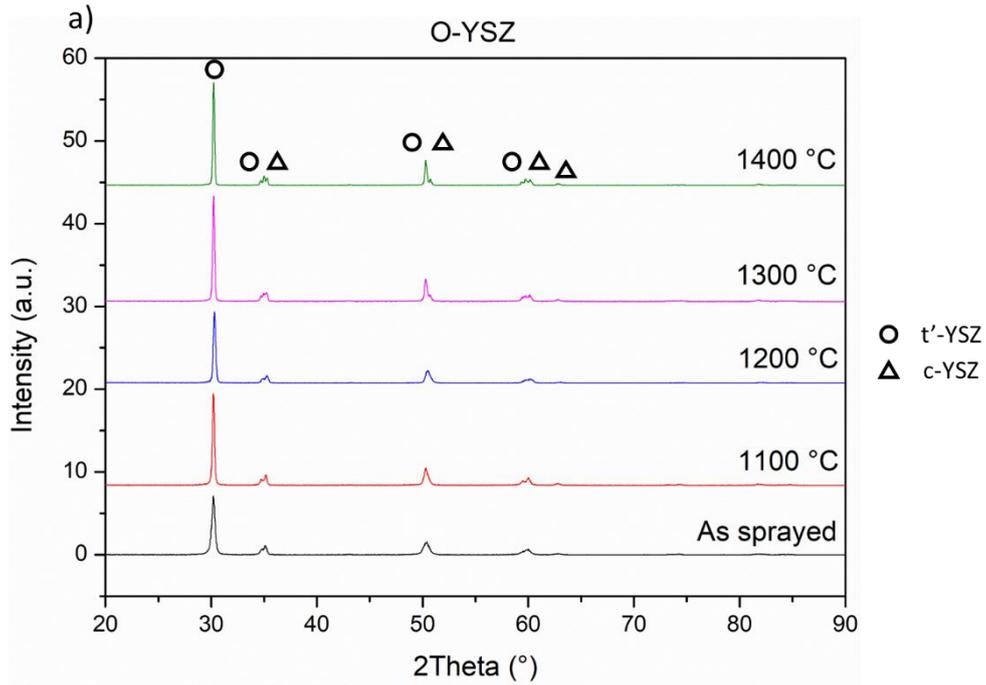
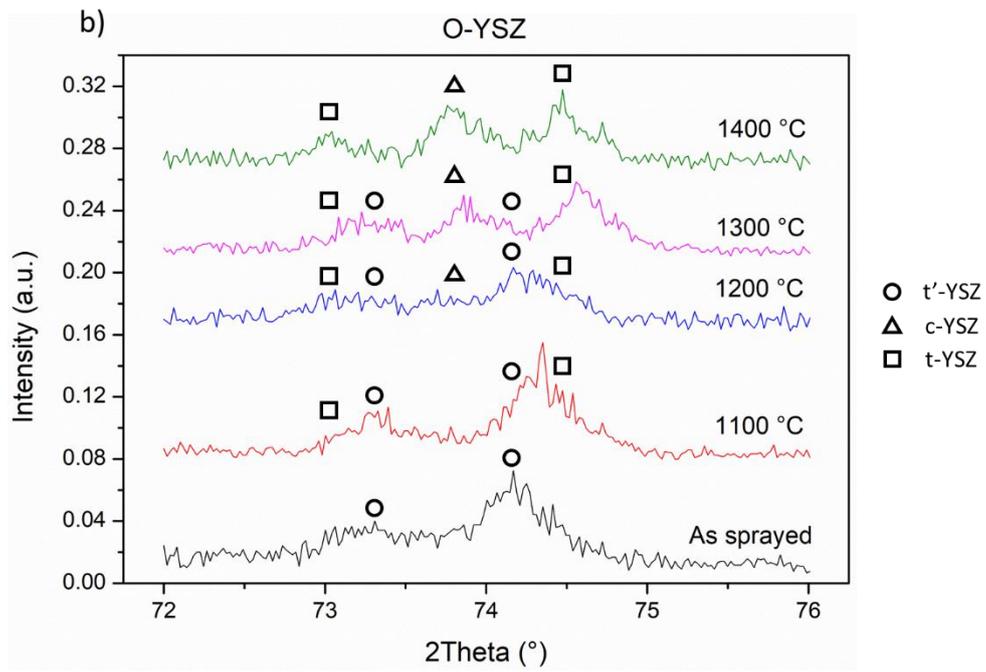

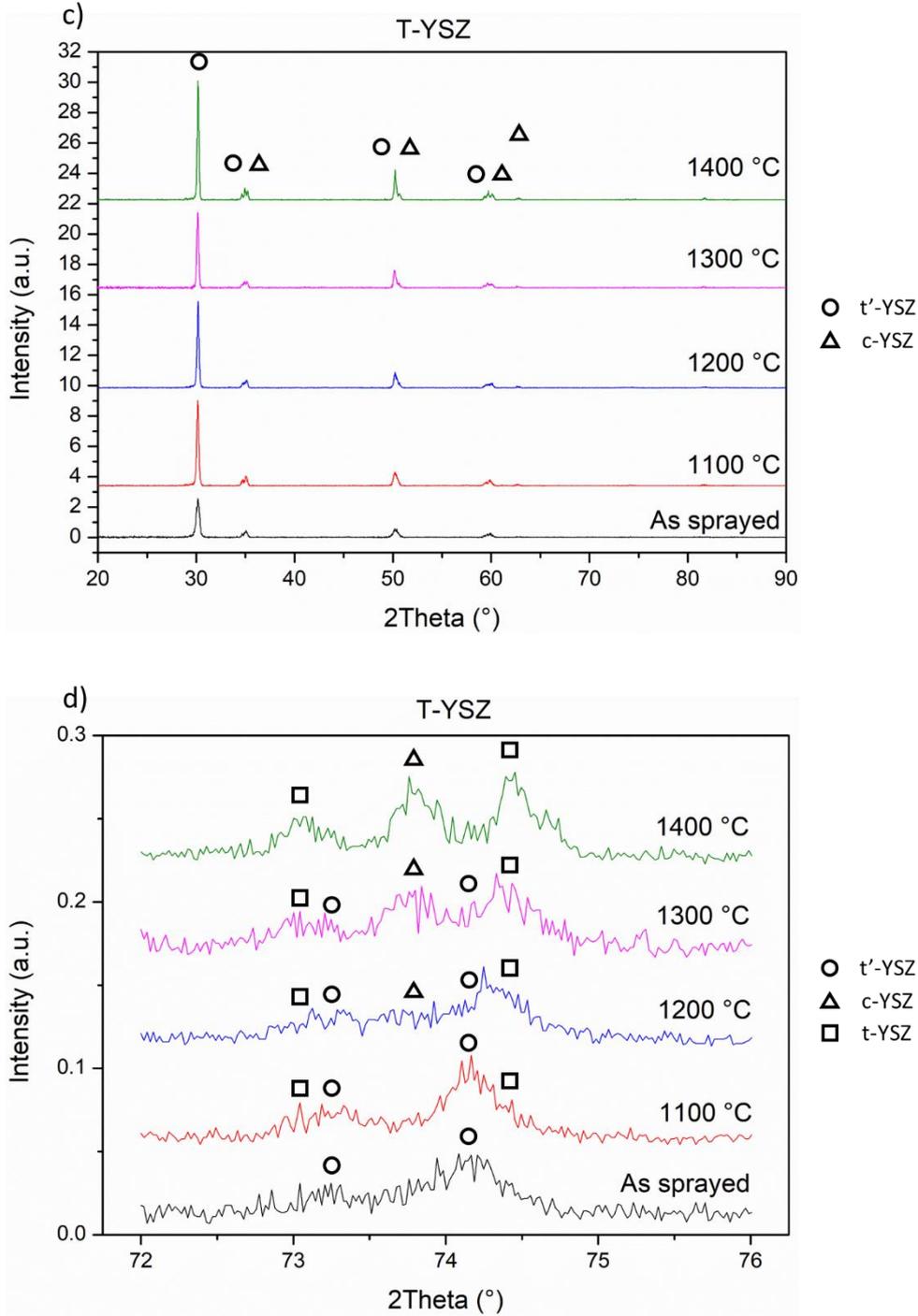

*Figure 5: XRD spectra for both O-YSZ (plots a and b) and T-YSZ (plots c and d) as-sprayed and heat-treated coatings. Plots a and c show the full spectra, whereas plots b and d show a detailed view of the range of interest for phase identification. The 72° - 76° scans show for both coatings that the transformation from the initial t' phase (circles) into a mixture of t (squares) and c (triangles) phases starts at 1100 °C and is only fully completed after 72 h at 1400 °C*

The data indicates that for both O-YSZ and T-YSZ samples, the phase composition of the as-sprayed coatings is only metastable tetragonal phase (i.e. *t'*-YSZ). Due to the close proximity between the main *t'*-YSZ peak (~74.1°) and the main *t*-YSZ peak (~74.4°), the determination of the phases is a challenging task. The data presented in Figure 5 seems to suggest that a small amount of decomposition is taking place at 1100 °C; however, since no evidence of *c*-YSZ peaks (~73.8°) could be identified, it could be due to the noise in the measurement. At 1200 °C both samples show clear signs of *t*-YSZ and *c*-YSZ peaks, although the decomposition is not completed yet, as the *t'*-YSZ peak is still quite predominant. A heat treatment at 1300 °C seems to mostly complete the *t'*-YSZ decomposition into *t*-YSZ and *c*-YSZ, with only traces of the initial *t'*-YSZ peak. The heat treatment at the higher temperature, 1400 °C, shows no evident signs of *t'*-YSZ phase, which would indicate that the decomposition is completed at or below this temperature. Even after 72 h at 1400 °C no evidence of monoclinic *m* phase could be detected.

To further investigate the effect that this phase transformation has on the heat-treated coatings, Rietveld refinement was conducted to calculate the tetragonality, as it is shown in Figure 6. Ilavsky *et al.* [20] has determined the cell parameter variations $c/a\sqrt{2}$ as a function of the amount of $YO_{1.5}$ in mol % (x) as the following equation over the concentration of $YO_{1.5}$ up to about 7 mol %, as shown in Equation 1 (data was retrieved from the original plot and linear fitted with $R^2 = 1$):

$$c/a\sqrt{2} = 1.02257 - 0.0032x \tag{1}$$

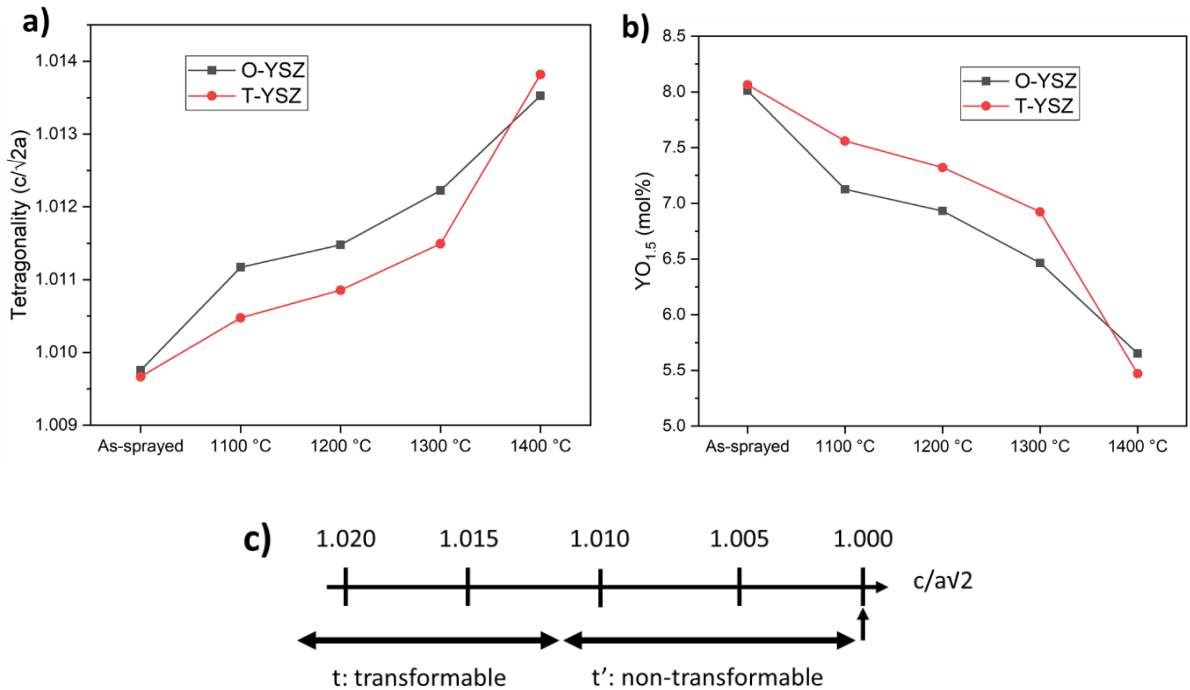

*Figure 6: a) Tetragonality and b) $YO_{1.5}$ composition (mol %) of O-YSZ and T-YSZ coatings as-sprayed and after various heat treatments, c) tetragonal forms of YSZ based on the tetragonality value [46]. Tetragonality increases in both coatings as the heat treatment temperature is increased, but T-YSZ presents a lower value, requiring a higher temperature to reach the threshold between transformable and non-transformable phases*

As it can be seen, the tetragonality of both as-sprayed coatings is approximately 1.0096, corresponding to the *t'* non-transformable phase region as indicated in Figure 6c. As the temperature of the heat treatment is increased, starting at 1100 °C, the tetragonality increases, reaching a value of ~1.011 in the case of O-YSZ. As the heat treatment temperature is increased, so does the tetragonality value, reaching the *t* transformable phase region, and leading to the appearance of the corresponding peaks in the XRD spectra presented in Figure 5. This same trend can be seen as the temperature is increased, with a noticeable increase in the tetragonality once the heat treatment temperature is 1400 °C. The same effect, although with the reduction of $YO_{1.5}$ composition can be seen in Figure 6, as expected. From the Rietveld refinement, it can also be calculated the crystallite size and the micro-strain of the coatings, as it is shown in Figure 7.

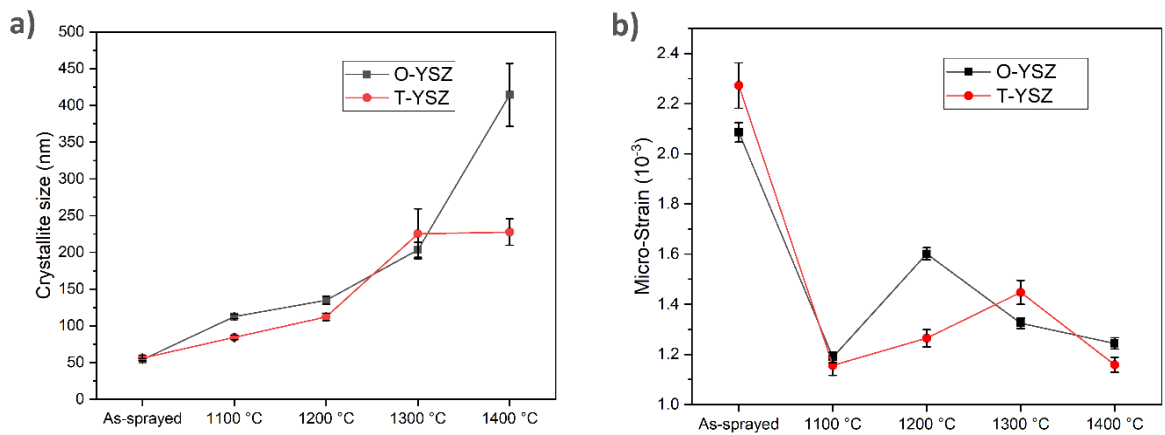

*Figure 7: a) Crystallite size and micro-strain for both O-YSZ and T-YSZ coatings on the as-sprayed condition and after various heat treatments. O-YSZ shows almost a two-fold increase in crystallite size at 1400 °C compared to T-YSZ. b) Micro-strain is reduced in both cases from the as-sprayed condition. O-YSZ presents a sharp peak at 1200 °C whereas T-YSZ shows a smaller peak, with a smoother transition, at 1300 °C*

The initial crystallite size for both coatings is approximately 50 nm, with a slow increase in crystallite size up to ~125 nm at 1200 °C. There is a further increase in size up to ~200 nm at 1300 °C for both coatings, but from this point the behaviour is different for each coating. In the case of O-YSZ, heat treatment at 1400 °C causes the crystallite size to increase up to ~400 nm, whereas for T-YSZ the same temperature produces no change with respect to 1300 °C. When considering the micro-strain, both coatings show a high level of micro-strain in the as-sprayed condition (~$2.2 \times 10^{-3}$), phenomenon expected on thermal sprayed coatings due to the rapid cooling experienced upon impact [47]. The micro-strain is reduced once the heat treatment at the lowest temperature, 1100 °C, is conducted (~$1.2 \times 10^{-3}$), suggesting an annealing-like process. From this point, the two coatings once more differ in their behaviour. The O-YSZ coating presents a micro-strain peak at 1200 °C, being the values at 1300 °C and 1400 °C lower. In the case of T-YSZ, the peak occurs at 1300 °C, and its magnitude is less than the one in O-YSZ. Both coatings have a low micro-strain value (~$1.2 \times 10^{-3}$) at 1400 °C.

As well as the phase content and porosity are related to the heat treatment temperature, such process has an impact on the mechanical properties of the coating, such as micro-hardness. Aiming to further understand the implications of the evolution of porosity within heat-treated 8YSZ coatings, the micro-hardness of the as-sprayed and heat-treated, free-standing coatings was measured. The relationship between micro-hardness and porosity, measured using both neutron scattering and IA, can be seen in Figure 8.

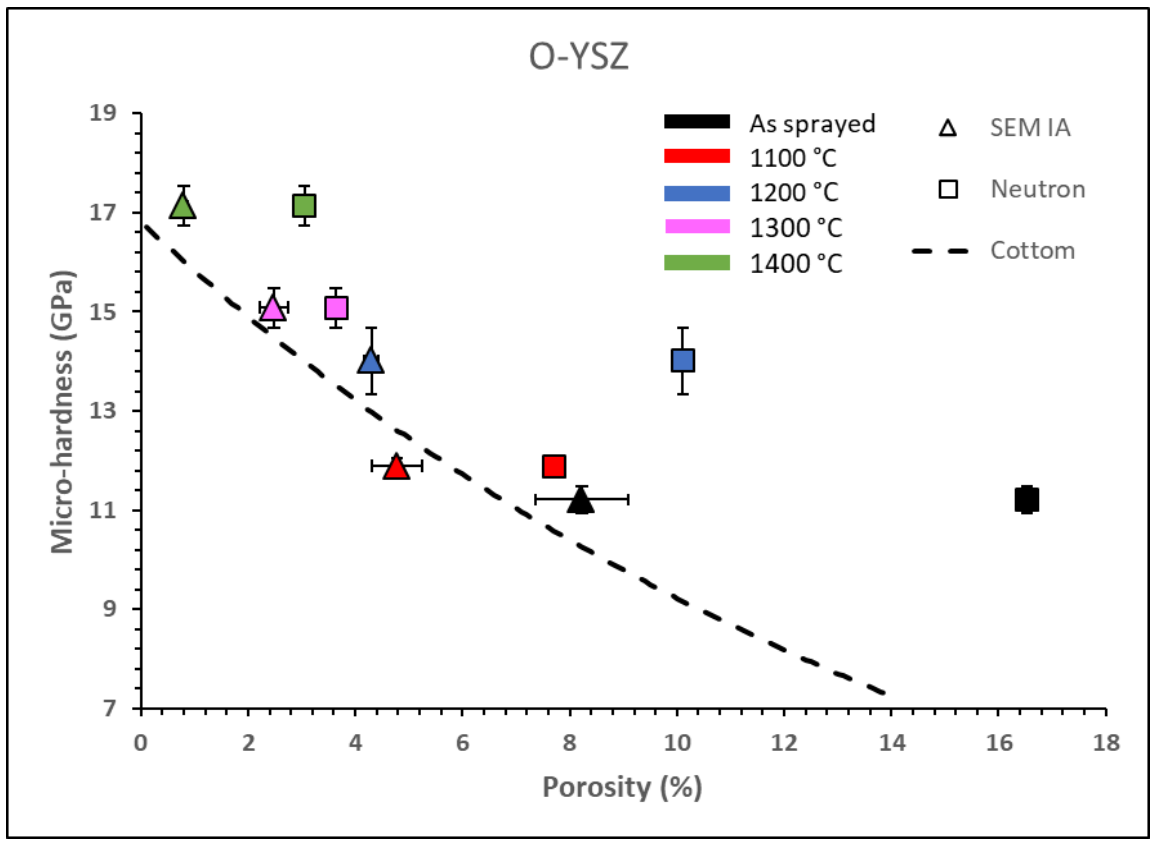

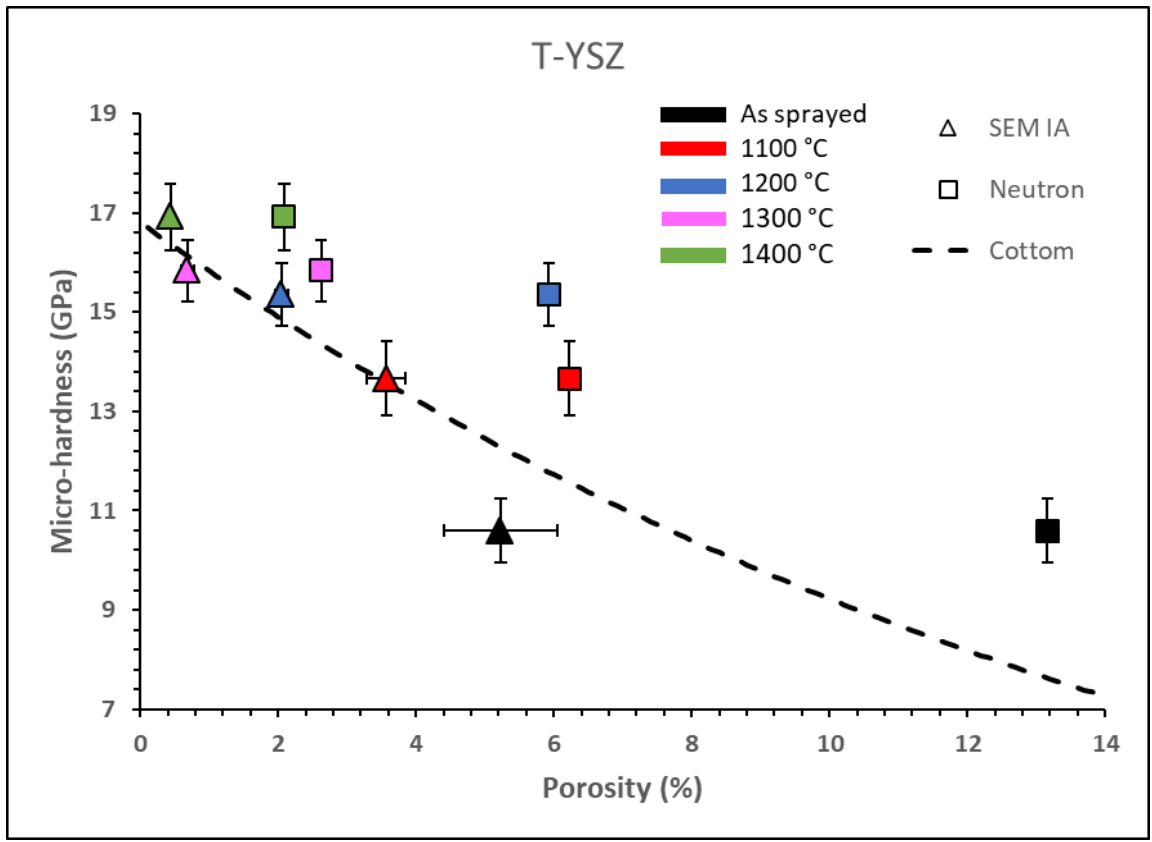

*Figure 8: Plot of the micro-hardness versus the measured porosity using neutron scattering (square) and image analysis (triangle). The dashed line corresponds to the empirical curve reported by Cottom et al. [48].*

In addition to the micro-hardness measurements, Figure 8 also shows a dashed line corresponding to the empirical curve reported by Cottom *et al.* [48], being derived from measurements of the porosity on sintered YSZ using IA. The results show a good agreement between the empirical curve and the measurements performed using IA. In the case of the neutron scattering data points, they all show a shift to the right, due to the higher porosity measured with this technique. There is also a similar trend in the relationship between micro-hardness and porosity, except for the points corresponding to the as-sprayed coating and 1200 °C, which will be further discussed in the next section.

## 4. Discussion

### 4.1. Porosity measurement: neutron scattering versus image analysis techniques

Both neutron scattering and IA techniques can be successfully applied to study the porosity present within thermal sprayed coatings, but some considerations should be taken first. Three main differences were observed in this work, namely, the pore size range measurable, the inclusion of micro-cracks as porosity features and the need for pore shape assumption.

First, as it was shown in Figure 1, the combination of SANS and USANS techniques allows the study of pores with a radius between ~1 nm and ~10 µm. In the case of IA, the lower limit for the radius was ~70 nm, with the largest feature observed being around ~700 nm. Due to the use of suspension HVOF thermal spraying as the deposition technique, pores with a radius below ~70 nm play a considerable role. Even after heat treatment at 1300 °C for 72 h, there is a considerable contribution of such porosity, which is not detectable using IA. This is one of the key factors contributing to the persistent higher value for the total porosity when measured through neutron scattering versus IA, particularly for heat treatment temperatures below 1300 °C, as it can be seen in Figure 4. In the comparison between SANS and IA reported by Deshpande *et al.* [49] the results consistently showed a porosity 1.25 times higher when measured using IA, being the opposite in this work. However, in their work, coatings were produced using powder-based thermal spray techniques. Secondly, micro-cracks influence the total porosity measured using neutron scattering, as this technique exposes the whole sample to the neutron beam, without the possibility of the user avoiding unwanted features. This would explain the sharp increase of porosity for the O-YSZ coating heat treated at 1200 °C observed in Figure 4. Neutron scattering provides a more complete picture, whereas IA allows for the determination of the amount of porosity and micro-cracks. Thirdly, as it was described in Appendix A, the shape of the pores had to be stablished for the analysis, which in this case corresponded to spheres. As it could be seen in the SEM-BSE cross-section images in Figure 2, this assumption was the less accurate for the as-sprayed samples. Given the large difference when compared to the IA porosity value, the initial hypothesis would be that neutron scattering techniques overestimate this value. This is further corroborated by the observed deviation from the empirical curve in the micro-hardness versus porosity plot (Figure 8). As the micro-hardness value is independent from the data modelling, to have the as-sprayed data point closer to the empirical relationship, the porosity value should be smaller. Therefore, it can be concluded that the porosity of the as-sprayed coatings is being overestimated when using neutron scattering techniques.

Regarding the differences between neutron scattering and IA size distributions, the first thing that should be considered is that only the range from 70 to 700 nm could be compared, although for neutron scattering there is much more information outside these limits. Secondly, the IA profile is obtained as the average of the five individual areas imaged. This has the effect of "smoothing out" outliers and individual features only present in one area. As a result, the black profiles in Figure 3 show the overall tendency of the pores with radius between 70 and 700 nm. In the case of neutron scattering, a single measurement per coating, which included the whole area submerged on the neutron beam, was performed. This allowed for a more detailed view of unique occurrences or populations of features. Such difference can be clearly seen in Figure 3a, b and c, showing the as-sprayed coating and heat treatments with temperatures up to 1200 °C. On these three instances, both techniques present a very similar picture, with differences arising as the heat treatment temperature is increased, promoting the appearance of micro-cracks, which increase the signal detected through neutron scattering. Despite the already mentioned difference experienced at 1200 °C between the porosity measurement for neutron scattering and IA, the size distribution shown in Figure 3c is still fairly consistent. This is due to the range investigated (70 to 700 nm), which doesn't includes the majority of the features observed in the neutron scattering volume distribution, as it can be seen in Figure 1. In the case of 1300 °C and 1400 °C the tendency changes, mainly due to detection of micro-cracks and similar features by the neutron scattering techniques, which translate into features with an associated radius values between 70 and 700 nm.

### 4.2. Porosity evolution

The porosity was studied attending to three main characteristics: pore shape, pore volume distribution and total porosity present in the coating. These three characteristics presented an evolution as the samples were heat treated. First, pore shape was highly irregular in the as-sprayed coatings, presenting a combination of elongated, inter-splat pores and globular pores, as represented in Figure 2. Heat treatment even at the lowest temperature (1100 °C) produced the consolidation of the porosity as globular pores, with only small traces of non-spherical pores. This effect could be seen to continue at higher heat treatment temperatures.

The porosity volume distribution obtained from the neutron scattering curves for both the 8YSZ coatings, presented in Figure 1, clearly shows how heat treatment at any temperature has a considerable impact on the porosity of the coatings. As it can be seen, porosity with a radius below 20 nm is greatly reduced at heat treatment temperatures as low as 1100 °C. If the heat treatment temperature is 1300 °C or above, this effect is extended to porosity with a radius below 100 nm, although to a lesser degree. One of the possible explanations for this very efficient reduction of porosity with a radius below 20 nm can be found when considering the crystallite size. The crystallite size of the as-sprayed coatings was determined to be ~50 nm for both coatings, as shown in Figure 7. During heat treatment, pores with a radii below 20 nm are reduced due to a densification process, which is favoured for pores smaller than the crystallite size [50]. In general, an increase in the heat treatment temperature had the effect of reducing the amount of fine pores in favour of porosity with a larger radius, producing an overall shift of the pore volume distribution to the right, as represented in Figure 1.

The effect on the total porosity is represented in Figure 4, with a reduction in the total porosity as the heat treatment temperature is increased. A different behaviour is observed at 1200 °C, with the porosity remaining the same, as in the T-YSZ coating, or even increasing as seen in the O-YSZ coating. A contribution to this phenomenon can be found in the SEM-BSE cross-section images taken during the study. As the image in Figure 9a shows, the O-YSZ coating heat treated at 1100 °C presented areas where the porosity was distributed along a line. Once the O-YSZ coating were heat treated above 1100 °C, these features were no longer present, an only micro-cracks (cracks with a length lower than 100 µm and not connected to the coating surface or the substrate interface) could be detected. This indicates that lined porosity acted as precursor of micro-cracks, as seen in Figure 9b.

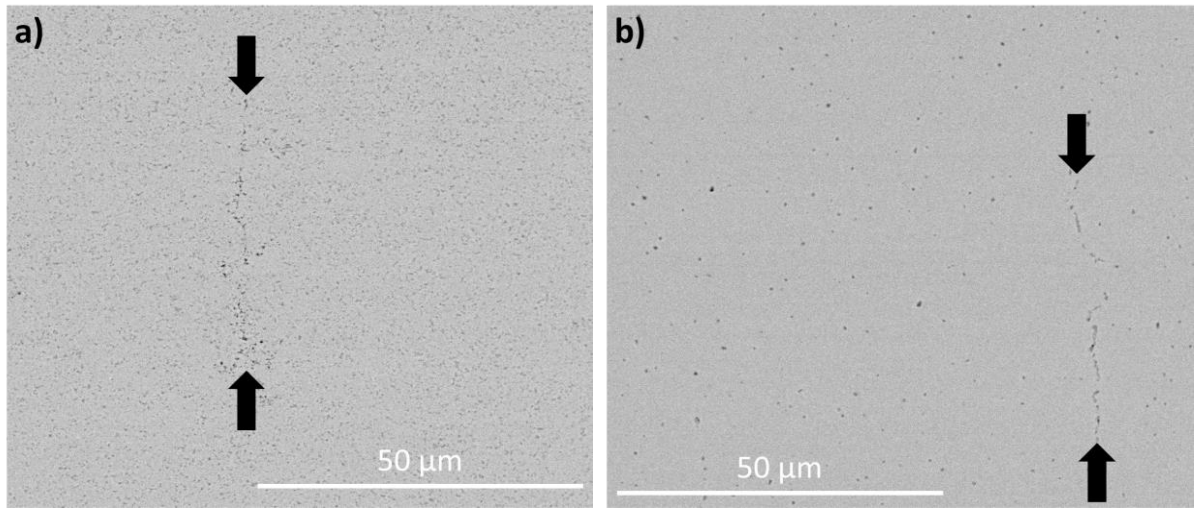

*Figure 9: SEM-BSE images of the cross-section of O-YSZ coating heat treated at a) 1100 °C and b) 1400 °C. The arrows mark the beginning and end of the lined porosity and micro-crack*

As discussed in section 4.1, micro-cracks are measured when using neutron scattering techniques, but not with IA, contributing to the difference between the two techniques at 1200 °C.

4.3. Phase transformation

Heat treatment will have an effect on the phase composition of the 8YSZ coatings, particularly as the temperature approaches 1200 °C, which is traditionally considered as the threshold for the decomposition of pure metastable tetragonal to tetragonal plus cubic phase, as discussed in section 3.2. As it could be appreciated in the XRD spectra in Figure 5, heat treatment at 1200 °C causes the initiation of the phase transformation from pure metastable tetragonal to tetragonal plus cubic phase. This temperature is consistent with the literature, with an established temperature of 1200 °C as the upper limit for the presence of metastable tetragonal phase. A shift in the metastable tetragonal phase peak at 2Θ ~74° is also observed, which has been associated with a decrease of the yttria content in the t phase and an increase in the c phase [20]. These phase transformation and yttria content reduction were confirmed through Rietveld refinement, as it can be seen in Figure 6 and Figure 7.

The tetragonality of both coatings is almost identical on the as-sprayed samples, but once the heat treatment at 1100 °C is conducted, O-YSZ presents a higher tetragonality value. At this temperature, the tetragonality value is ~1.011, almost at the transition point between the initial *t'* non-transformable phase region and the *t* transformable phase region, as reported by Ilavsky *et al.* [20]. As the heat treatment temperature is increased to 1200 °C, there is a further reduction in the $YO_{1.5}$ composition, accompanied by another increase in the tetragonality. This additional increase is enough to enable the phase decomposition from metastable tetragonal to stable tetragonal and cubic phase, although only in the case of O-YSZ. The tetragonality value for T-YSZ at 1200 °C is still below 1.011. This threshold value is passed at 1300 °C in the case of T-YSZ. This phase transformation has implications regarding the micro-strains present within the coating. As observed by Witz *et al.* [44], the micro-strain present on thermal-sprayed YSZ coatings after heat treatment at 1100 °C for 100 h is ~0.4%; however, the micro-strain on the coating after heat treatment at 1200 °C for 100 h was measured to be ~0.7%, due to the reduction in the metastable tetragonal phase content. This increase in the micro-strain as the phase decomposition takes place can be seen in Figure 7. In the case of O-YSZ, as mentioned before, there is a sharp increase at 1200 °C, coinciding with the transition from *t'* non-transformable to *t* transformable, whereas for T-YSZ the peak is located at 1300 °C. Both the sharpness and the maximum magnitude of the peak are less than for O-YSZ.

This phase transformation is believed to be the cause for the appearance of the previously mentioned micro-cracks. In addition, the presence of aligned porosity, as it could be seen in Figure 9a, acts as preferential sites for micro-cracks to form once the thermal stresses accumulate within the coating. Such micro-cracks would correspond to the features observed in the scattering profiles above 100 nm and at 800 – 900 nm. The reason why this phenomenon has a higher effect on the O-YSZ coating could be due to the differences observed in the micro-strain profile in the T-YSZ, reaching at 1300 °C a lower micro-strain value (~1.4 × $10^{-3}$) than the O-YSZ coating at 1200 °C (~1.6 × $10^{-3}$). In addition, O-YSZ suspension showed a larger particle size distribution, as well as a higher content of larger pores in the as-sprayed sample, when compared to T-YSZ. This increase in pores of larger size could lead to more aligned pores, producing micro-crack initiators.

Once the heat treatment temperature is further increased to 1300 °C or 1400 °C, the phase transformation from pure metastable tetragonal to tetragonal plus cubic phases seems to be mostly completed. The appearance of the cubic phase at this temperature produced a reduction of ~1.7% in the micro-strain of the coating. In addition, the higher temperatures, once they had completed the phase transformation, produced a relaxation of the accumulated micro-strain. These phenomena contributed to the self-healing of micro-cracks to some degree, as it was appreciated during SEM imaging (not shown here). This would explain why the disproportionate increase in porosity, which has been associated with micro-cracking, was only seen at 1200 °C. Above this temperature, micro-cracking still took place, explaining why the porosity is still higher than the IA value, but to a lesser degree than at 1200 °C. In no case presence of monoclinic phase could be detected, which indicates that in both coatings a heat treatment of 1400 °C for 72 h is not enough to promote the transformation from tetragonal to monoclinic phase. This corroborates the results reported by Ilavsky *et al.* [20] where plasma sprayed YSZ coating heat treated at 1400 °C for 100 h did not show appreciable traces of monoclinic phase. Presence of monoclinic phase was detected by Witz *et al.* [44] after 1000 h at 1300 °C, which indicates that longer exposures at lower temperature can cause the formation of monoclinic phase. A general schematic showing the evolution observed in 8YSZ coatings during heat treatment up to 1400 °C for 72 h is presented in Figure 10.

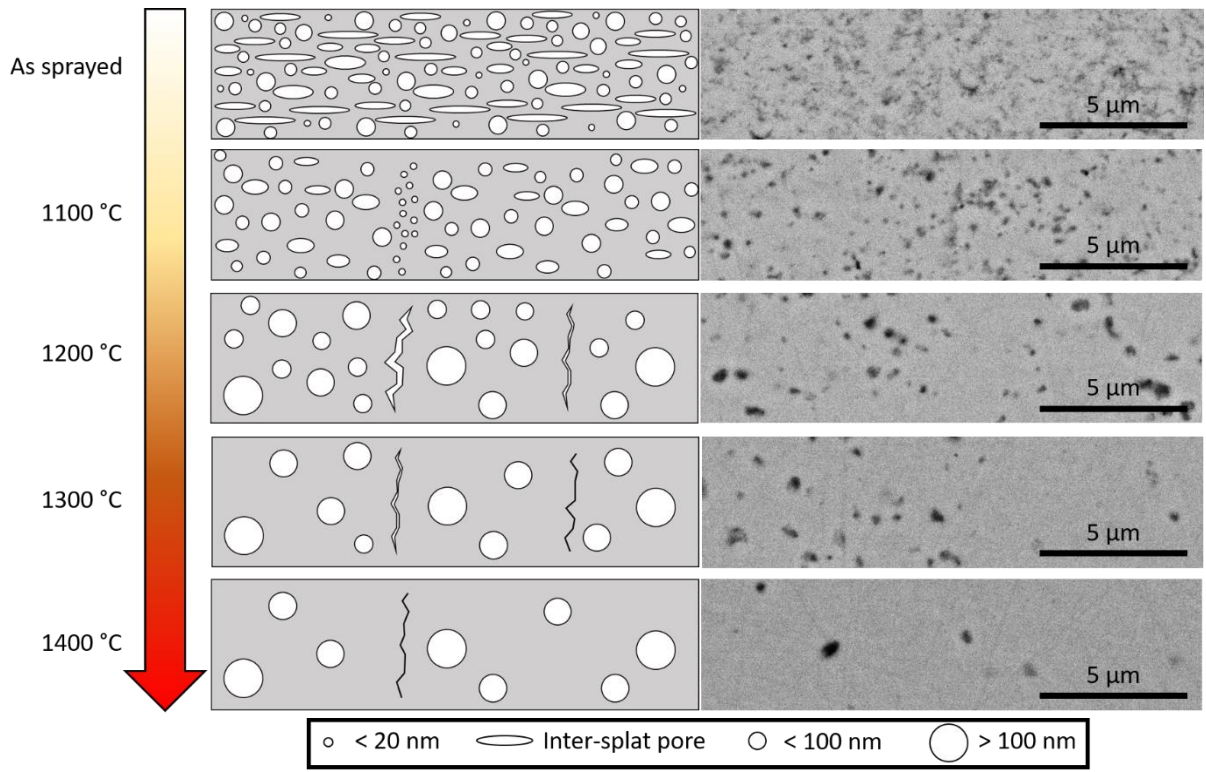

*Figure 10: Schematic representation of the evolution of porosity on thermal sprayed 8YSZ coatings (left) along with the corresponding BSE images of the cross-section of the coatings (right)*

The schematic show in Figure 10 represents the evolution of porosity on thermal sprayed 8YSZ coatings, at the as-sprayed condition and after heat treatment at several temperatures for 72 h. The 1100 °C heat treatment reduces porosity below 20 nm, almost all non-globular porosity and aligned porosity can be seen. Heat treatment at 1200 °C further reduces porosity below 100 nm and shows micro-cracks due to the phase transformation and accumulation of thermal stresses. Higher temperatures produce a reduction of the finer pores and some degree of self-healing of the micro-cracks

## 5. Conclusions

In this work, we have demonstrated that 8YSZ coatings deposited using SHVOF thermal spray present a large amount of nano-sized porosity, contrary to more traditional powder-based thermal spray techniques such as APS, and a mixture of elongated, inter-splat pores and spherical pores. SANS & USANS and IA techniques were used to investigate the evolution of porosity after heat treatment at 1100 °C, 1200 °C, 1300 °C and 1400 °C for 72 h. Both techniques were capable of measuring the pore size distribution and total porosity; however, for suspension thermal sprayed coatings, due to the presence of nano-sized pores, neutron scattering is needed to study the entire range of porosity present. IA is needed to visually inspect the microstructure of the porosity, assessing artefacts and shape, information required for an accurate modelling of the SANS and USANS data.

The results show that heat treatment at 1100 °C reduces the presence of nano-pores and overall porosity, with a transition from elongated pores into spherical pores. Heat treatment at 1200 °C induced, in addition to a continuation of the evolution of the microstructure, a phase transformation from the original metastable tetragonal into stable tetragonal and cubic phases. This process, which was accompanied by a raise in the micro-strain, led to the formation of micro-cracks within the coating. Heat treatments at higher temperatures (1300 and 1400 °C) completed the phase transformation, without the formation of monoclinic phase, limiting the accumulation of thermal stresses, further reducing the presence of fines pores and presenting a degree of self-healing of micro-cracks.

**Appendix A**

The SANS technique is based on the different scattering length densities (SLD) present within a heterogeneous material, being in this case assumed that the pores have SLD equal to zero, and the 8YSZ coating has a SLD of 5.4 × $10^{-10}$ $cm^{-2}$. The contrast in SLD between the coating and the pores gives rise to the coherent elastic scattering of the monochromatic neutron beam, which can be detected and quantified. A general schematic of the scattering process is shown in Figure A.1.

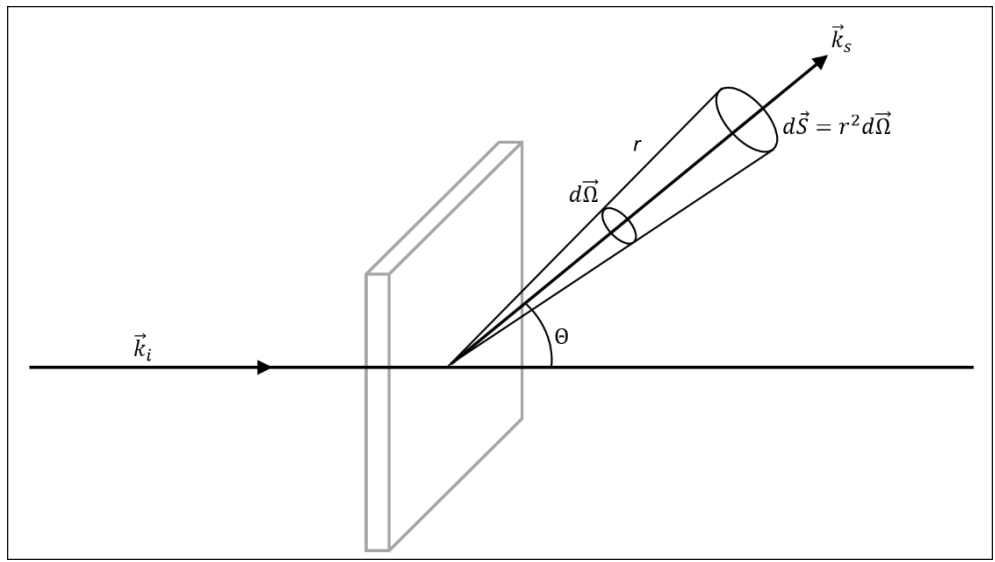

Figure A.1: Schematic representation of the neutron scattering process of an incident neutron beam ($\vec{k}_i$) with a rectangular sample, producing a scattered neutron beam ($\vec{k}_s$) with an angle θ inside a solid angle $d\vec{\Omega}$, resulting in a microscopic differential scattering cross-section $d\vec{S}$

In the case of the USANS technique, the core concept remains the same as with SANS, being the scattering of a monochromatic neutron beam due to difference on the SLD of the material. Nevertheless, in order to be able to detect lower scattering angles (corresponding to larger physical features in the sample) smaller wavelengths and shorter flight paths are used. Two Si (311) single crystals are placed in front and behind the sample, using the diffraction of the neutrons to highly collimate the beam, creating a strong correlation between scattering angles and wavelengths. This improves the flux on the sample while allowing detailed angular resolution.

The differential scattering cross-section can be defined as described in Equation A.1.

$$\frac{d\Sigma}{d\Omega}(Q) = \frac{number\ of\ neutrons\ scattered\ per\ second\ into\ d\Omega}{\Phi d\Omega} \quad (A.1)$$

Where $\vec{Q}$ is the scattering vector (or momentum transfer) $\vec{k_s} - \vec{k_i}$ with magnitude $Q = (4\pi/\lambda)sin\Theta$ and being $\Phi$ the flux. The differential scattering cross-section has units of $cm^2$. This is what the detectors pick up during the experiment, being necessary a reduction step, where the data is normalised over the sample volume. After said step the normalised differential scattering cross-section *I(Q)* is obtained, with units of $cm^{-1}$.

To model the scattering data collected, *I(Q)*, two factors are used: a form factor (which in this work has been chosen to be a sphere) and a size distribution (for the analysis a normal Gaussian distribution was used). The form factor of a sphere can be expressed as shown in Equation A.2.

$$I_{Sphere}(Q, R, \Delta\eta) = \left[\frac{4}{3}\pi R^3 \Delta\eta 3 \frac{sinQR - QRcosQR}{(QR)^3}\right]^2 \quad (A.2)$$

Whereas the Gaussian size distribution is characterised by Equation A.3.

$$Gauss(x, N, s, x_0) = \frac{N}{c} e^{-\frac{(x-x_0)^2}{2s^2}} \quad (A.3)$$

Where N is a scaling factor of the size distribution, x is the radius of the pores, $x_0$ is the mean of the distribution and s is the standard deviation of the distribution. c is chosen so that $\int Gauss(x, s, x_0)dR = N$. Therefore, the Gaussian size distribution can be rewritten as $Gauss = N \times p(x)$ where $\int p(x)dx = 1$. The combination of both the form factor and the size distribution can be used as a model for the analysis of the neutron scattering data, as shown in Equation A.4.

$$I_{model}(Q) = \int_0^\infty Np(R)I_{Sphere}dR \rightarrow I_{model}(Q) = N\int_0^\infty I_{Sphere}dR \quad (A.4)$$

In Equation A.4, *N* has an additional physical meaning, being the number density of scatterers per unit volume. Therefore, this value can be used to both calculate the volume distribution as $N(R) \times R^3$, and the volume fraction of spheres within the sample volume, as detailed in Equation A.5.

$$f_p = \int_0^\infty n(R)\frac{4}{3}\pi R^3 dR = N\frac{4}{3}\pi \langle R^3 \rangle \quad (A.5)$$

The modelling was done using the SASfit software [51], version 0.94.10. It follows a non-linear least square fitting routine. Several populations of spheres were added, each one of them following the I$_{model}$(Q) equation described above, until the combined signal matched the measured data. Once the fitting was performed, the values of $N(R)$ and $\langle R^3 \rangle$ were exported from SASfit to calculate the volume distribution and volume fraction for each population of spheres. The complete curve of the volume distribution was constructed following the convolution of the individual curves from each population of spheres, while the total volume fraction was calculated as the sum of each volume fractions, multiplied by 100 to express the value as a percentage.

## 6. Acknowledgments

This work was supported by the Engineering and Physical Sciences Research Council (EPSRC) (grant number EP/L016206/1) as well as by an Innovation Placement from the EPSRC Centre for Doctoral Training in Innovative Metal Processing (IMPaCT). The authors would like to thank John Kirk for his assistance during the SHVOF thermal spray and Dr. Acacio Rincon Romero for the fruitful discussions regarding densification of YSZ.